\documentstyle[preprint,aps]{revtex}

\def\journal#1, #2, #3, 1#4#5#6{
    {\sl #1~}{\bf #2}, #3 (1#4#5#6)}

\def\prd{\journal Phys. Rev. D, }
\def\prl{\journal Phys. Rev. Lett., }

\def\cmp{\journal Comm. Math. Phys., }
\def\np{\journal Nucl. Phys., }
\def\pl{\journal Phys. Lett., }
\def\annp{\journal Ann. Phys. (N.Y.), }

\def\Tr{\hbox{Tr}}
\def\half{{1\over2}}
\def\podd{\pi^{\rm odd} }

\def\LL{{\cal L}}
\def\a{\alpha}
\def\la{\lambda}
\def\Ga{\Gamma}
\def\Godd{\Gamma^{odd}}
\def\eps{\epsilon}
\def\mnl{{\mu\nu\la}}
\def\dab{\delta_{ab}}
\def\eab{\eps_{ab}}
\def\beq{\begin{equation}}
\def\eeq{\end{equation}}

\begin{document}
\preprint{UdeM-GPP-TH-94-5, hep-th/9408091}

\title{On the Coleman-Hill Theorem}

\author{Avinash Khare}
\address{Institute of Physics, Sachivalaya Marg, Bhubaneswar-751005, India}

\author{R. MacKenzie and M.B. Paranjape}
\address{Laboratoire de physique nucl\'eaire, Universit\'e de
Montr\'eal C.P. 6128, succ. centreville, Montr\'eal, Qu\'ebec, Canada, H3C 3J7}
\maketitle

\begin{abstract}
\widetext
The Coleman-Hill theorem prohibits the appearance
of radiative corrections to the topological mass (more precisely, to
the parity-odd part of the vacuum polarization tensor at zero momentum)
in a wide class of
abelian gauge theories in 2+1 dimensions. We re-express the theorem
in terms of the
effective action rather than in terms of the vacuum polarization tensor.
The theorem so restated becomes somewhat stronger: a known
exception to the theorem, spontaneously broken scalar Chern-Simons
electrodynamics, obeys the new non-renormalization theorem. Whereas the
vacuum polarization {\sl does} receive a one-loop, parity-odd correction, this
does not translate to a radiative contribution to the Chern-Simons term in the
effective action. We also point out a new situation, involving scalar
fields and parity-odd couplings, which was overlooked in the original
analysis, where the
conditions of the theorem are satisfied and where the topological mass
does, in fact, get a radiative correction.
\narrowtext
\end{abstract}

The existence of the Chern-Simons (CS) term in 2+1 dimensional gauge
theories\cite{cs} has fueled a large body of research over the last several
years, in fields varying from condensed matter physics to pure mathematics.
The term leads to fractional-statistics excitations (relevant to the
fractional quantum Hall effect)\cite{fqhe}, while its topological nature in
the nonabelian case has yielded information on the classification of
lower-dimensional manifolds and knot invariants \cite{math}. It is odd
under parity, and provides for a gauge-invariant mass for the relevant
vector bosons.

The coefficient of the non-abelian CS term must be quantized for the theory
to be consistent\cite{desjactem,redlich,polychronakos}.
This quantization must be
respected by radiative corrections, and, indeed, in pure $SU(N)$ gauge
theory, it has been found that the coefficient (appropriately normalized so
that
the quantization is to integer values) receives a one-loop correction which
changes its value by the integer $N$\cite{pisrao}.

If the gauge field is coupled to matter fields which spontaneously break
the symmetry, the situation is much more delicate in the non-abelian case.
With complete breaking of the symmetry, the topological mass itself
receives a correction which is a complicated function of the parameters of
the theory, and certainly no quantization condition is satisfied, in
general \cite{khlsch}. However, the quantization of the
coefficient of the CS term
itself might be salvaged, since there exist other terms which are not of a
topological nature and therefore whose coefficients
need not be quantized, yet which contribute to the topological mass. The
non-quantization of the radiative correction to the topological mass might
thus be a combination of a quantized correction to the CS term along the
lines of \cite{pisrao}\ along with a non-quantized correction to the other
terms, as was suggested \cite{khlsch}.

More alarming is the case of a non-abelian theory spontaneously broken to a
non-abelian subgroup. There, the topological mass is found {\it not to be}
quantized (similar to the situation in \cite{khlsch}), yet there are no
terms other than the CS term which make a contribution to the topological
mass \cite{khamacpanpar}.
It would appear, therefore, that in such theories the CS
term does receive a non-quantized radiative correction, and thus that they
are not consistent at a quantum level, following the reasoning of
\cite{desjactem,redlich,polychronakos}.

A parallel but quite different situation arises in the abelian case, where
no quantization condition is required. There is thus no {\it a prioro}
restriction on radiative corrections, yet such corrections are in fact few
and far between. Coupling the photon to a fermion yields a correction to
the linear term in a momentum expansion of the parity-odd part of the photon
vacuum polarization tensor  $\podd_{\mu\nu}$
(whether or not the CS term is there
initially) at one loop\cite{cs,desjactem,oneloopfermion}, but not at two
loops\cite{twoloopfermion}. Inspired by this unexpected result, Coleman and
Hill\cite{colhil} devised a proof that, under very general conditions, the only
correction to the linear term in a momentum expansion of $\podd_{\mu\nu}$
comes
from fermions at one loop. In particular, they have emphasized that the
result is valid even for nonrenormalizable interactions in the presence of
gauge- and Lorentz-invariant regularization.

Situations exist where the conditions given by Coleman and
Hill are satisfied,
yet where radiative corrections to the topological mass do nonetheless
arise. Namely, this can occur
if there are new parity-violating interaction
terms in the initial Lagrangian, a possibility which was overlooked in
\cite{colhil}. For
instance, if the photon is
coupled to massive vector particles which themselves violate parity (a
possibility in 2+1 dimensions) there is a correction to the topological
mass \cite{hagpanram}.

Furthermore, even scalar fields can have
such parity-violating interactions. Indeed, the following interaction
Lagrangian
\beq
\LL_{int}=j_\mu\left(ieA^\mu+{\alpha\over e}\epsilon^{\mu\nu\lambda}
F_{\nu\lambda}\right)
\label{preone}
\eeq
yields, after a straightforward evaluation of the vacuum polarization,
\beq
\podd_{\mu\nu}\left.{\eps^{\mnl}p_\lambda\over2p^2}\right|_{p^2=0}
\equiv\podd(0)={\alpha m\over\pi},
\label{stuff}
\eeq
where $m$ is the mass of the scalar and $j^\mu$ is the usual particle
current.

In addition to these actual counter-examples to the Coleman-Hill theorem,
a number of situations have been found where the initial
assumptions of the theorem are not satisfied, and where the
vacuum polarization tensor does get further radiative corrections. One such
situation is if there are massless particles present, in which case
infrared divergences spoil the proof of the
theorem\cite{semsodwu-,slava,chen}. Another is if Lorentz or gauge
invariance is not manifest, a situation found in the nonabelian case
(where, as outlined above, radiative corrections indeed exist).

A third such case is that of spontaneously broken scalar
electrodynamics\cite{khlebnikov,spiridonov}, where the interaction
term explicitly violates one of Coleman and
Hill's initial assumptions. (In this case, in their words, there are parts
of the gauge boson kinetic terms ``lurking'' about in the interaction
Lagrangian.) Here, it has been found that, with even
an infinitesimal CS term at the tree level, a macroscopic parity-odd part of
the vacuum polarization is induced to one loop. Since the tree
level Lagrangian respects parity (in the absence of a CS term), this system
exhibits spontaneous parity violation, very reminiscent of the case of massless
spinor electrodynamics\cite{oneloopfermion}. This induced parity violation is
not found in the phase where the symmetry is respected; indeed, the limit of
letting the bare CS term go to zero does not commute with that of letting the
expectation value of the scalar field go to zero.

In the case of spontaneous symmetry breaking, particularly,
the relation between
the parity-odd part of the vacuum polarization and the renormalization of the
CS term is rather indirect. This is best seen within the framework of the
effective action. The vacuum polarization tensor, in a phase with
scalar field expectation value $\phi_0$,
is the second derivative of the effective
action with respect to the photon, evaluated at zero photon field and at
$\phi=\phi_0$:
\beq
\pi_{\mu\nu}(x,y)=\left.\left\{{\delta^2\Gamma_{\rm eff}
\over\delta A^\mu(x)\delta A^\nu(y)}\right\}\right|_{A=0,\phi=\phi_0}.
\label{one}
\eeq
We are interested in the parity-odd part of this for small
momenta. By Lorentz invariance, this will be proportional to
$\eps_{\mu\nu\rho}k^\rho$; let us call the coefficient $\podd(0)$.

$\podd(0)$ will certainly receive contributions from the tree-level CS term
as well as from radiative corrections to it. But, as we will show, other terms
in the effective action which reduce to the CS term if one sets $\phi\to\phi_0$
(which we refer to as ``would-be CS terms'') will also contribute to
$\podd(0)$,
making the extraction of the radiative correction to the CS term itself more
complicated. In fact, if we rephrase the statement of the Coleman-Hill theorem
in terms of the non-renormalization of the coefficient of the CS term in the
effective action, then, as will be shown below,
at least in the case of minimally-coupled scalar electrodynamics, the
newly-stated
theorem remains valid even in the presence of spontaneous symmetry breaking.
More precisely, in spontaneously-broken, minimally-coupled scalar
electrodynamics, the coefficient of the CS term in the effective action
{\it does not receive} a radiative correction at one loop.

To be specific, the model we consider consists of a real scalar doublet with
gauged SO(2) symmetry and with bare CS term, described by the Lagrangian
\beq
\begin{array} {l}
\LL=-{1\over4}{F_{\mu\nu}}^2+{\mu\over2}\eps^{\mnl}A_\mu\partial_\nu A_\la
-{1\over2\xi}(\partial_\mu A^\mu)^2
\\
\qquad +\half{(D_\mu\phi)_a}^2
-{m^2\over2}\phi^2-{\la\over4!}\phi^4-{\tau\over6!}\phi^6,
\label{two}
\end{array}
\eeq
where $\phi^2=(\phi_a)^2$ and
$(D_\mu\phi)_a=(\partial_\mu\dab-eA_\mu\eab)\phi_b$. The third term is
for gauge fixing; we will eventually work in the Landau gauge
$\xi\to0$.

In what follows, we will compute the terms of interest in the effective action
$\Ga[\hat\phi,\hat A]$ to one loop, the ultimate goal being to compute the
one-loop correction to the CS term, and to show that it is indeed zero.
Naively, the CS term is calculated by computing the term in $\Ga$ which is
parity-odd, bilinear in $\hat A$, and which contains one derivative. In
the phase of unbroken symmetry, this is perfectly unambiguous and
correct; however, in the presence of spontaneous symmetry breaking there are
would-be CS terms which have the identical structure if we let $\phi\to\phi_0$,
and which necessitate additional work in order to separate them from the true
CS
term.

To
see this, consider the low-momentum terms in the parity-odd part of $\Ga$. Such
terms must be expressed in terms of $\eps_{\mnl}$; one finds
\beq
\begin{array} {l}
\Godd[\hat\phi(x),\hat A(x)]=\int d^3x\,\eps_\mnl
\bigl(c_1 \hat A_\mu\partial_\nu \hat A_\la
+c_2(\hat\phi^2)\hat\phi_a D_\mu\hat\phi_a\partial_\nu \hat A_\la
\\ \qquad\qquad\qquad\qquad
+c_3(\hat\phi^2)\eab\hat\phi_a D_\mu\hat\phi_b\partial_\nu \hat A_\la
+\hbox{higher order terms}\bigr).
\label{three}
\end{array}
\eeq
We can, in fact, put the coefficient $c_2$ to zero without loss of
generality, since $c_2(\hat\phi^2)\hat\phi_a D_\mu\hat\phi_a$ is a total
derivative, and integration by parts demonstrates that the term itself is,
in fact, also a total derivative.

If we were to proceed according to the naive approach outlined above, setting
$\hat\phi\to\phi_0$, the third term would make an unwanted
contribution to the effective
action:
\beq
\Godd[\phi_0,\hat A(x)]=\int d^3x\,\eps_\mnl
\left(c_1 +e \phi_0^2 c_3(\phi_0^2)\right)\hat A_\mu\partial_\nu \hat
A_\la+\cdots.
\label{four}
\eeq
We must therefore perform a supplementary calculation in order to separate the
undesired $c_3$ part from the genuine CS term.

This can be most easily done by
considering a field configuration whose scalar part has a space-dependent
piece:
$\hat\phi=\phi_0+\phi_1(x)$. If we now evaluate $\Godd$ to linear order in
$\phi_1$ and in $\hat A$, the CS term makes no contribution, and we find
\beq
\Godd[\phi_0+\phi_1(x),\hat A(x)]=\int d^3x\,\eps_\mnl
c_3\eab\phi_{0a}\partial_\mu\phi_{1b}\partial_\nu\hat A_\la+\cdots.
\label{five}
\eeq
This second calculation then gives $c_3$, which
then enables us to extract $c_1$.

To zero loops, the effective action is the
ordinary action: $\Ga_0[\hat\phi,\hat A]=S[\hat\phi,\hat A]=\int d^3x\,
\LL[\hat\phi,\hat A]$. The one-loop contribution to the effective action is
obtained according to the following prescription\cite{jackiw}. One expands the
ordinary action about the desired field values, $S[\hat\phi+\phi,\hat A+A]$.
It is the part quadratic in $A$ and $\phi$
which is relevant to one loop; this is
\beq
\begin{array} {l}
S^q=\int d^3x\,\Biggl\{\half A_\mu\left(
g_{\mu\nu}\partial^2-\left(1-{1\over\xi}\right)\partial_\mu\partial_\nu
+\mu\eps_{\mu\a\nu}\partial_\a+e^2\hat\phi^2
g_{\mu\nu}\right)A_\nu
\\ \qquad\qquad
+A_\mu\left(e\eps_{ac}\left(\hat\phi_c\partial_\mu
-(\partial_\mu\hat\phi_c)\right)
+2e^2\hat A_\mu\hat\phi_a\right)\phi_a
\\ \qquad\qquad
+\half\phi_a\biggl(-\left(\partial_\mu\dab-e\hat A_\mu\eab\right)
\left(\partial_\mu\delta_{bc}-e\hat A_\mu\eps_{bc}\right)
\\ \qquad\qquad\quad
-\left(m^2+{\la\over6}\hat\phi^2+{\tau\over120}\hat\phi^4\right)\delta_{ac}
-\left({\la\over3}+{\tau\over30}\hat\phi^2\right)\hat\phi_a\hat\phi_c\biggr)
\phi_c\Biggr\}
\\ \quad\ \,
\equiv\int d^3x\,\Bigl\{\half A_\mu U_{\mu\nu}(\hat\phi)A_\nu
+A_\mu V_{\mu a}(\hat\phi,\hat A)\phi_a
\\ \qquad\qquad
+\phi_a\half W_{ab}(\hat\phi,\hat A)\phi_b\Bigr\}
\label{six}
\end{array}
\eeq
The one-loop contribution to the effective action $\Ga_1$ is then obtained by
functional integration; the result is
\beq
\Ga_1[\hat\phi,\hat A]={i\over2}\Tr\log W
+{i\over2}\Tr\log(U-V W^{-1} V).
\label{eight}
\eeq

Were we interested in the effective potential,
we could take constant field values and evaluate the traces exactly. However,
for terms in the effective action involving derivatives,we must allow
$\hat\phi$
and $\hat A$ to depend on $x$, and some sort of approximation must be employed.
Fortunately, we are interested only in the CS and would-be CS terms, so an
expansion in derivatives of the fields and in powers of the fields themselves
will suffice.

Let us first outline the calculation of the combined genuine and would-be CS
terms, letting $\hat\phi=\phi_0$.  We must compare (\ref{four}),
on the one hand, with an expansion of (\ref{eight}) on the
other.\footnote{For simplicity, we redefine $c_1$ so that it excludes
the original (zero-loop) coefficient of the CS term, $\mu/2$.}
The first term
in (\ref{eight}) can be ignored since it makes no contribution to $\Godd$. In
the second term, the argument of the log can be written
\beq
X\equiv U-V W^{-1}
V=U^{(0)}-(V^{(0)}+V^{(1)})(W^{(0)}+W^{(1)}+W^{(2)})^{-1}(V^{(0)}+V^{(1)}),
\label{nine}
\eeq
where the superscripts indicate powers of $\hat A$. We can expand $X$ in powers
of $\hat A$, $X=X^{(0)}+X^{(1)}+X^{(2)}+\cdots$; higher terms are unnecessary
since we are only interested in terms quadratic in $\hat A$. The relevant terms
in the one-loop, parity-odd part of the effective action are
\beq
\Godd_1[\phi_0,\hat A]={i\over2}\Tr[((X^{(0)})^{-1})^{odd}X^{(2)}]
-{i\over2}\Tr[((X^{(0)})^{-1})^{odd}X^{(1)}((X^{(0)})^{-1})^{even}X^{(1)}]
+\cdots.
\label{ten}
\eeq
There are rather a large number of terms; however, after some work, only
one
survives; re-expressing it in terms of $U$, $V$ and $W$,
\beq
\Godd_1[\phi_0,\hat A]=-{i\over2}\Tr\left[
{\left(U^{(0)}-V^{(0)}{W^{(0)}}^{-1}V^{(0)}\right)^{-1}}^{odd}_{\mu\nu}
V^{(1)}_{\nu a}{W^{(0)}}^{-1}_{ab}V^{(1)}_{\mu b}\right]+\cdots.
\label{eleven}
\eeq
The relevant expressions can be read more or less directly off (\ref{six}). The
trace is not yet calculable, since it involves both derivatives (in the
zeroeth-order parts) and space dependance (in the first-order parts). A
separation of these can be achieved in a derivative expansion \cite{aitfra},
which suffices for our purposes; the trace can then be performed. After some
work, one finds (after Wick rotation)
\beq
\Godd_1[\phi_0,\hat A(x)]={4\mu e^4\over3}\int d^3x\,\eps_{\mnl}\phi_0^2
\hat A_\mu\partial_\nu\hat A_\la\int{d^3p\over(2\pi)^3}
{p^2\over\left( (p^2+e^2\phi_0^2)^2 +\mu^2p^2\right)
(p^2+m_L^2)^2}+\cdots,
\label{twelve}
\eeq
where $m_L^2=m^2+\la\phi_0^2/5+\tau\phi_0^4/24$ and the integral is in momentum
space; comparing with (\ref{four}), the coefficients of the CS and would-be CS
terms satisfy the following relation:
\beq
c_1 +e \phi_0^2 c_3(\phi_0^2)={4\mu e^4\phi_0^2\over3} I,
\label{thirteen}
\eeq
where $I$ is the integral in (\ref{twelve}); it is easy to evaluate,
although not particularly transparent.

To calculate $c_3$, we proceed in a similar fashion, this time giving a
space-dependent piece to the scalar field: $\hat\phi=\phi_0+\phi_1(x)$.  The
term linear in $\phi_1$ and in $\hat A$ is, after some work,
\beq
\Godd_1[\phi_0+\phi_1(x),\hat A(x)]=
{4\mu e^3\over3}\int d^3x\,\eps_{\mnl}\phi_0\eab\partial_\mu\phi_{1b}
\partial_\nu\hat A_\la\, I+\cdots.
\label{fourteen}
\eeq

Combining (\ref{five}), (\ref{thirteen}) and
(\ref{fourteen}), we find our main result: the radiative correction to the
coefficient of the CS term in the effective action is
\beq
c_1=0.
\label{fifteen}
\eeq

In summary, we have shown that in the Higgs phase of 2+1 dimensional scalar
electrodynamics, all one-loop contributions to the topological mass arise
from manifestly gauge invariant terms in the effective action which reduce
to the CS term once the scalar field is set equal to its vacuum expectation
value, rather than being attributable to the CS term itself. This suggests
a more general theorem: namely, that the Coleman-Hill theorem restated as a
non-renormalization theorem for the coefficient of the CS term in the
effective action would be valid. We have also shown, in passing, that
one-loop corrections to the topological mass arise simply from
parity-violating interactions, which exist even for scalar fields, in
addition to the known cases of fermions \cite{desjactem} and vector bosons
\cite{pisrao,hagpanram}.

These ideas fit in nicely with the suggestion of Khlebnikov and
Schaposhnikov \cite{khlsch} for the case of a completely spontaneously
broken non-abelian gauge symmetry, where an apparent violation of the
quantization condition on the topological mass was postulated to be
attributable to the existence of other would-be CS terms. There, however,
the analysis is much more involved and has yet to be done. The case of
partial breaking to a non-abelian subgroup remains a puzzle
\cite{khamacpanpar}: the violation of the quantization condition on
the topological mass has thus far eluded a similar explanation, since no
would-be CS terms can be constructed in an analogous fashion.

As this manuscript was being finalized, a paper has appeared which
discusses the renormalization of the CS term in self-dual and
supersymmetric versions of the Abelian Higgs model \cite{kaoleeleelee}.

A.~Khare thanks the Laboratoire de Physique Nucl\'eaire for the kind
invitation and hospitality during his visit.
This work was supported in part by NSERC of Canada and by FCAR du Qu\'ebec.

\end{document}